\documentclass{article}
\usepackage{amsmath}
\usepackage{amssymb}
\usepackage{graphicx}
\begin{document}
\begin{center}
\LARGE
\textbf{Quantum Probability's Algebraic Origin}
\\[1 cm]
\normalsize
Gerd Niestegge
\\[0,3 cm]
\footnotesize
gerd.niestegge@web.de
\\[1 cm]
\end{center}
\normalsize
\begin{abstract}
Max Born's statistical interpretation made probabilities play a major role in quantum theory. 
Here we show that these quantum probabilities and the classical probabilities have very different origins. While the latter always result from an assumed probability measure, the first include transition probabilities with a purely algebraic origin. Moreover, the general definition of transition probability introduced here comprises not only the well-known quantum mechanical transition probabilities between pure states or wave functions, but further physically meaningful and experimentally verifiable novel cases.

A transition probability that differs from 0 and 1 manifests the typical quantum indeterminacy in a similar way as Heisenberg's and others' uncertainty relations
and, furthermore, rules out deterministic states
in the same way as the Bell-Kochen-Specker theorem.
However, the transition probability defined here 
achieves a lot more beyond that: it demonstrates that 
the algebraic structure 
of the Hilbert space quantum logic 
dictates the precise values of certain 
probabilities and it provides an unexpected access to these quantum probabilities 
that does not rely on states or wave functions.
\\[0,5 cm]
\end{abstract}

\section{Introduction}

The Boolean algebra (or the equivalent Boolean lattice) is a mathematical structure playing an important role in many scientific and technical fields such as logic, probability theory, circuitry, computer science. Only quantum theory challenges the general applicability of this structure, since the dichotomic observables (those with spectrum $\left\{0,1\right\}$) do not form a Boolean algebra, but a lattice where the distributivity law fails \cite{beltrametti1984logic, birkhoff-vN36, varadarajan1968and1970}. 

The system of the dichotomic observables is called the \emph{quantum logic} and becomes the framework for a new general definition of the \emph{transition probability}. This definition includes not only the well-known quantum mechanical transition probabilities between pure states or wave functions, but further physically meaningful and experimentally verifiable novel cases. It is pointed out that the transition probabilities have a purely algebraic origin, which has mostly been ignored in the past. In this way, they become very different from the classical probabilities which result from probability measures.

Other approaches to the transition probabilities are possible; one is based on projective quantum measurement (L\"uders - von Neumann quantum measurement process) \cite{Nie1998HPA01} and another is based on a non-Boolean extension of the conditional probabilities \cite{niestegge2001non}. The approach presented here, based on the new definition, is more elementary than these two since it does not require advanced concepts like quantum measurement or the non-Boolean conditional probabilities. Some basic knowledge of quantum mechanics and linear algebra (here particularly the Cauch-Schwarz inequality) are enough to understand the paper.

After a brief sketch of the quantum logic and its state space in section 2, 
the transition probability will be defined formally in section 3. In section 4, 
some examples will be studied to reveal the link to the 
well-known quantum mechanical transition probabilities 
and to identify some further novel cases. The connection to 
the typical quantum indeterminacy 
is discussed in section 5.

\section{Quantum logic and states}

Commonly, the quantum mechanical observables are mathematically represented 
by self-adjoint (Hermitian) linear operators on a Hilbert space $H$. The dichotomic observables 
(those with spectrum $\left\{0,1\right\}$) become self-adjoint projection operators 
and form the \emph{quantum logic} $L_H$. It includes $0$ and the identity~$\mathbb{I}$.
By considering the one-to-one relation between the self-adjoint projection operators
and the closed linear subspaces of $H$, it becomes evident that $L_H$ is a lattice 
with order relation $\leq$, infimum $\wedge$ and supremum $\vee$. Moreover, $0$ is 
the smallest element, $\mathbb{I}$ is the largest element in $L_H$ and, 
for any $p \in L_H$, $p' := \mathbb{I} - p$ is the orthogonal complement of $p$.

For $p,q \in L_H$, $p \leq q$ is equivalent 
to each one of the following conditions:
$pq = p$, $qp = p$, $pqp = p$ or $qpq = p$.
A pair $p,q \in L_H$ is called \emph{orthogonal} if $p \leq q'$ or
if one of the following equivalent conditions holds: 
$q \leq p'$, $pq=0$, $qp = 0$, $pqp=0$ or $qpq=0$. 
Orthogonality means that $p$ and $q$ mutually exclude each other.

A \emph{state} shall allocate a probability to each element of the quantum logic 
in a consistent way and thus becomes a map 
$\mu : L_H \rightarrow \left[0,1\right] := \left\{s \in \mathbb{R} : 0 \leq s \leq 1 \right\}$ 
with $\mu(\mathbb{I})=1$ and
$\mu(p+q) = \mu(p) + \mu(q)$
for each orthogonal pair $p,q \in L_H$ \cite{beltrametti1984logic}.
The states form a convex set $S(L_H)$ which is called the \emph{state space}.

If the dimension of the Hilbert space is two, it is necessary to distinguish 
between $S(L_H)$ and the subset $S_{lin}(L_H)$; $S_{lin}(L_H)$ consists of those states that 
can be extended to a linear map defined for all bounded observables. If it exists,
this linear extension is unique because of the spectral theorem and is denoted by $\mu$ again.
Due to Gleason's theorem \cite{christensen1982measures, gleason1957measures}, the identity 
$S(L_H) = S_{lin}(L_H)$ holds for all other Hilbert space dimensions ($\neq 2$).
While $S_{lin}(L_H)$ is associated with the linear structure of the observables, 
$S(L_H)$ depends on the algebraic structure of the quantum logic only.

\section{Transition probability}

The novel definition of the transition probability
shall now be presented. If a pair $p,q \in L_H$ with $p \neq 0$ and some $r \in [0,1]$
satisfy the identity 
\begin{center}
$\mu(q) = r$ for all $\mu \in S_{lin}(L_H)$ with $\mu(p)=1$,
\end{center}
$r$ is called the \emph{transition probability from} $p$ \emph{to} $q$ 
and is denoted by $\mathbb{P}(q|p)$. The identity
$\mathbb{P}(q|p) = r$ then becomes equivalent to the set inclusion
\begin{equation*}
\left\{\mu \in S_{lin}(L_H) : \mu(p)=1\right\} \subseteq \left\{\mu \in S_{lin}(L_H) : \mu(q)=r\right\}
\end{equation*}
and means that, whenever the probability of $p$ is $1$, 
the probability of $q$ is determined and must be $r$.
Particularly in the situation after a quantum measurement
that has provided the outcome $p$, the probability of $q$ 
becomes $r$, independently of the initial state 
before the measurement.

If $0 \neq p_2 \leq p_1$ and $\mathbb{P}(q|p_1)$ exists with $p_1,p_2,q \in L_H$, 
then $\mathbb{P}(q|p_2)$ exists and $\mathbb{P}(q|p_1) = \mathbb{P}(q|p_2)$.
This follows immediately from the above definition.

The transition probability $\mathbb{P}(q|p)$ is a characteristic 
of the algebraic structure of the observables.
If the Hilbert space dimension does not equal two, we have $S(L_H) = S_{lin}(L_H)$ and 
the transition probability becomes a characteristic of the even more basic structure 
of the quantum logic.
\\[0,3 cm]
\textbf{Theorem.} 
\itshape
Suppose that $p \neq 0$ and $q$ are elements in the quantum logic $L_H$. 
\begin{enumerate}
\item[(i)]
The transition probability from $p$ to $q$ exists and
$\mathbb{P}(q|p) = r$ iff the linear operators $p$ and $q$ satisfy 
the simple algebraic identity 
$$pqp = rp.$$
\item[(ii)]
If $p \neq 0 \neq q$ holds and if both transition probabilities 
$\mathbb{P}(q|p)$ and $\mathbb{P}(p|q)$ exist, they are equal: 
$$\mathbb{P}(q|p) = \mathbb{P}(p|q).$$
\end{enumerate}
\normalfont
\textbf{Proof.}
(i) The linear extension of a state $\mu$ becomes a positive linear functional
and the Cauchy-Schwarz inequality holds:
$$\left|\mu(xy)\right| \leq \left(\mu(x^{2}\right)^{1/2} \left(\mu(y^{2}\right)^{1/2}$$
for all bounded linear operators $x$ and $y$ \cite{sakai1971}.
This implies that, for $p \in L_H$ with $\mu(p)=0$,
$\mu(xp) = \mu(px) = 0$ for all bounded linear operators $x$. 
Note that projections are idempotent ($p^{2} = p$).

$\Leftarrow$: Suppose $pqp = rp$. If $\mu \in S_{lin}(L_H)$ and $\mu(p)=1$, then $\mu(p')=0$ and 
$\mu(q) = \mu(pqp) + \mu(p'qp) + \mu(pqp') + \mu(p'qp') = \mu(pqp) = r \mu(p) = r$.
Therefore, $\mathbb{P}(q|p) = r$.
\newpage
$\Rightarrow$: Now suppose $\mathbb{P}(q|p) = r$. Let $\mu$ be any state in $S_{lin}(L_H)$.
If $\mu(p) = 0$, then $\mu(pqp) = 0 = \mu(rp)$.
If $\mu(p) > 0$, define a state $\mu_p \in S_{lin}(L_H)$ by 
$\mu_p(x)=\mu(pxp)/\mu(p)$ for $x \in L_H$. Then $\mu_p(p) = 1$ and therefore
$r = \mu_p(q) = \mu(pqp)/\mu(p)$. We now have $\mu(pqp) = r \mu(p)$ for all $\mu \in S_{lin}(L_H)$
and thus $pqp = rp$. 

(ii) Suppose that $p \neq 0 \neq q$ holds and that 
$\mathbb{P}(q|p)$ and $\mathbb{P}(p|q)$ both exist.
By (i) we have that $pqp = r_1 p$ and $qpq = r_2 q$ with 
$r_1 = \mathbb{P}(q|p)$ and $r_2 = \mathbb{P}(p|q)$.
Then $r_1 pq = pqpq = r_2 pq$ and either $r_1 = r_2$ 
or $pq = 0$. In the second case, $pqp = 0 = qpq$ and 
therefore $r_1 = 0 = r_2$. 
\hfill $\square$
\\[0,3 cm]
An immediate consequence of the theorem is that
$\mathbb{P}(q|p) = 1 $ iff $p \leq q$ and that 
$\mathbb{P}(q|p) = 0 $ iff $p$ and $q$ are orthogonal.

The transition probability is invariant under unitary transformations $u$: 
$\mathbb{P}(q|p)$ exists iff $\mathbb{P}(uqu^{-1}|upu^{-1})$ exists, and 
$\mathbb{P}(q|p) =\mathbb{P}(uqu^{-1}|upu^{-1})$. 
This follows from the above theorem and also directly from the
definition of the transition probability.

If the transition probability $\mathbb{P}(q|p)$ exists, one can 
use the \emph{trace} (\textit{tr}) to calculate it:
$tr(pq) = tr(p^{2}q) = tr(pqp) = \mathbb{P}(q|p) tr(p)$
and therefore $$\mathbb{P}(q|p) = tr(pq)/tr(p).$$
The term $tr(pq)/tr(p)$ always exists (unless $p=0$), 
but it represents a transition probability as defined above 
only in certain cases. The trace and the last equation 
cannot help to identify these cases.

There is an interesting connection between the transition probability defined here
and the \emph{equiangularity} studied in \cite{Freedman_2019}: if the projections $p$ and $q$ 
are equiangular, the transition probabilities $\mathbb{P}(q|p)$ 
and $\mathbb{P}(p|q)$ both exist. Transition 
probabilities are not considered in \cite{Freedman_2019}, but this follows by 
combining theorem 2.3 from there with the above theorem.

\section{Examples}

\textbf{Example 1.} \textit{Suppose that $p \neq 0$ and $q$ commute and that $\mathbb{P}(q|p) = r$ exists. 
Then $pqp = rp$ and $rpq = (pq)^{2} = pq$ and either $pq=0$ or $r = 1$. 
In the first case, $p$ and $q$ are orthogonal and $\mathbb{P}(q|p) = 0$.
In the second case, we have $p \leq q$ and $\mathbb{P}(q|p) = 1$.
This means that a non-trivial transition probability ($0 < \mathbb{P}(q|p) < 1$)
requires that $p$ and $q$ do not commute.}
\\[0,3 cm]
\textbf{Example 2.} \textit{Suppose that $\left|\psi\right\rangle$ is 
a normalized element of the Hilbert space $H$.
With the common bra-ket notation (Dirac notation), 
$p := \left|\psi\right\rangle \left\langle \psi \right|$ becomes the projector on the 
one-dimensional subspace generated by $\left|\psi\right\rangle$. For any other ${q \in L_H}$
we then have
$pqp = \left|\psi\right\rangle \left\langle \psi \right| q \left|\psi\right\rangle \left\langle \psi \right| = \left\langle \psi \right| q \left|\psi\right\rangle p$. This means that $\mathbb{P}(q|p)$
exists for all elements $q \in L_H$ with $\mathbb{P}(q|p) = \left\langle \psi \right| q \left|\psi\right\rangle$. The map ${q \rightarrow \mathbb{P}(q|p)}$ becomes the pure state defined by $\left|\psi\right\rangle$.}
\newpage
\noindent
\textbf{Example 3.} 
\itshape
If $q$, too, is the projector on a one-dimensional 
subspace and if this subspace is generated 
by the normalized element $\left|\phi\right\rangle \in L_H$, we have 
$q := \left|\phi\right\rangle \left\langle \phi \right|$ and 
$$\mathbb{P}(q|p) = \left| \left\langle \psi | \phi\right\rangle \right| ^{2}.$$ 
This is the well-known quantum mechanical transition probability 
between the Hilbert space elements 
$\left|\psi\right\rangle$ and $\left|\phi\right\rangle$, 
often known as wave functions or pure states.
\normalfont
\\[0,3 cm]
However, the general and abstract definition of the transition probability in section 3
goes beyond this situation. The following example demonstrates that the existence of
$\mathbb{P}(q|p)$ does not require $p$ to be a projector on a one-dimensional space. 

A non-zero transition probability 
$\mathbb{P}(q|p) = r \neq 0$
requires that the dimension of the image $qH$ of $q$ is not smaller than 
the dimension of the image $pH$ of $p$. This can be seen in the following way:
the identity $pqp = rp$ with $r \neq 0$ implies that $pqp$ and $p$ 
have the same image $pH = pqpH \subseteq pqH$;
therefore $dim (pH) \leq dim (pqH) \leq dim (qH)$.

If $0 < \mathbb{P}(q|p) < 1$, then $\mathbb{P}(q'|p) = 1 - \mathbb{P}(q|p) \neq 0$
and the dimensions of the images of both $q$ and its orthogonal complement $q'$
cannot be smaller than the dimension of the image of $p$. Therefore, 
a case with $0 < \mathbb{P}(q|p) < 1$ and $dim(pH) > 1$ requires that 
the dimension of the Hilbert space $H$ is not less than four. 
\\[0,3 cm]
\textbf{Example 4.}
\itshape
Consider the matrices
\begin{center}
$p := \left(
\begin{array}{cccc}
 1 & 0 & 0 & 0 \\
 0 & 1 & 0 & 0 \\
 0 & 0 & 0 & 0 \\
 0 & 0 & 0 & 0 \\
\end{array}
\right)
$ and 
$q := \left(
\begin{array}{cccc}
 {s_1}^{2} + {s_2}^{2} & 0                     & s_1 s_3   & - s_2 s_3 \\
 0                     & {s_1}^{2} + {s_2}^{2} & s_2 s_3   & s_1 s_3 \\
 s_1 s_3               & s_2 s_3               & {s_3}^{2} & 0 \\
 - s_2 s_3             & s_1 s_3               & 0         & {s_3}^{2} \\
\end{array}
\right)$
\end{center}
with $s_1,s_2,s_3 \in \mathbb{R}$ and ${s_1}^{2} + {s_2}^{2} + {s_3}^{2} = 1$.
Some matrix calculations show that $p^{2}=p$, $q^{2}=q$ (i.e. $p,q \in L_H$) 
and that $pqp = \left({s_1}^{2} + {s_2}^{2}\right) p$. The theorem then yields that 
$\mathbb{P}(q|p)$ exists with 
$$\mathbb{P}(q|p) = {s_1}^{2} + {s_2}^{2} = 1 - {s_3}^{2}.$$
\normalfont

Many further examples can be constructed 
by using $upu^{-1}$ and $uqu^{-1}$ instead of $p$ and $q$
with any unitary transformation $u$; 
then $\mathbb{P}(uqu^{-1}|upu^{-1}) = \mathbb{P}(q|p)$.

Only if $p$ is a projector on a one-dimensional space,
the transition probability $\mathbb{P}(q|p)$ can be represented
in the familiar way with the inner product of the Hilbert space 
as in the examples 2 and 3.
In the general case, however, 
this is not possible; nevertheless, $\mathbb{P}(q|p)$ constitutes 
the physically meaningful and experimentally verifiable 
probability of $q$ in the situation after a quantum measurement
that has provided the outcome $q$. The initial state before 
the measurement becomes irrelevant here in the same way as with 
a projector $p$ on a one-dimensional space.
\newpage
\noindent
\textbf{Example 5.} 
\textit{With the same $p$ as in the last example, $\mathbb{P}(q|p)$
exists only for some, but not for all $q \in L_H$. 
Since we have $dim(pH) = 2$, the above dimension considerations 
in connection with example  4 show that 
$\mathbb{P}(q|p)$ cannot exist for any $q$ that is a projector
on a one-dimensional subspace and not orthogonal to~$p$.}

\textit{If $dim(pH) \neq 1$, $\mathbb{P}(q|p)$ exists only for some, 
but not for all $q \in L_H$ and, therefore, 
the definition of a state by $q \rightarrow \mathbb{P}(q|p)$ fails.}
\\[0,3 cm]
\textbf{Example 6.}
\textit{A case, where $\mathbb{P}(q|p)$ exists, but $\mathbb{P}(p|q)$ does not
exist, can be constructed by using any projector $p$ on a one-dimensional space 
and any projector $q$ with $dim(qH) \geq 2$ and $pq \neq 0$ 
(i.e. $p$ and $q$ are not orthogonal). Since $dim(pH) = 1$, $\mathbb{P}(q|p)$
exists, and since $dim(pH) < dim(qH)$, $\mathbb{P}(p|q)$ cannot exist.}

\section{Quantum indeterminacy}

Our common sense, philosophy, logic and the classical sciences make us think 
that each proposition is either true or false; there is nothing in between. 
Is it possible to allocate an 
attribute `true' or `false' in a consistent way to each element of the quantum logic $L_H$?
Replacing `true' by $1$ and `false' by $0$, this would result in a state $\mu$ with 
$\mu(p) \in \left\{0,1\right\}$ for all $p \in L_H$. Such a state is 
called \emph{deterministic} (or \emph{dispersion-free}).
However, the Bell-Kochen-Specker theorem tells us that a deterministic state is impossible on
the Hilbert space quantum logic $L_H$ except that the dimension 
of the Hilbert space is two \cite{Bell1966, Cabello_1996, KocSp1967}.

This can also be seen by considering the 
transition probabilities.
Suppose that $0 < \mathbb{P}(q|p) < 1$ holds for $p,q \in L_H$ and 
that $\mu$ is a deterministic state.
Since $\mu(p)=1$ would imply $\mu(q) = \mathbb{P}(q|p) \notin \left\{0,1\right\}$,
it follows that $\mu(p)=0$. We thus get $\mu(p)=0$ for 
all projectors $p$ on one-dimensional subspaces and, furthermore, for 
all orthogonal sums of such projectors. With a 
finite-dimensional Hilbert space $H$, this would exhaust all elements of the 
quantum logic $L_H$ including $\mathbb{I}$ and yield a contradiction 
to $\mu(\mathbb{I})=1$. Therefore, $S_{lin}(L_H)$ does not include 
any deterministic state and, for $dim(H) \neq 2$, there is no 
deterministic state at all.

However, the transition probability $\mathbb{P}(q|p)$ achieves more 
than just ruling out determinism. It dictates the precise value 
of the probability of $q$,
whenever $p$ carries the probability 1 
(particularly in the situation after a quantum measurement
that has provided the outcome $q$), 
and is a characteristic of the algebraic structure of the quantum logic. 

The most famous manifestation of the typical quantum indeterminacy 
is Heisenberg's \emph{uncertainty relation} \cite{heisenberg1927}. 
Further more general uncertainty relations are due to 
Robertson \cite{Robertson1929} and Schr\"odinger \cite{Schroedinger1930}.
A transition probability $\mathbb{P}(q|p) = r$ with $0 < r < 1$ also represents 
a kind of uncertainty relation: if $p$ is known with certainty,
$q$ must be unknown and carry the probability $r$.
\newpage
\section{Conclusions}

It is well-established that the algebraic structure 
of the Hilbert space quantum logic rules out deterministic states
(Bell-Kochen-Specker theorem \cite{Bell1966, Cabello_1996, KocSp1967}). 
In the present paper, it has been seen that the algebraic structure 
does a lot more beyond that; it dictates the precise values of the 
transition probabilities which thus provide an 
unexpected access to quantum probability 
that does not rely on states or wave functions.

If $p \in L_H$ and $q \in L_H$ commute, only three cases are possible for the transition 
probability: either it does not exist or $\mathbb{P}(q|p) = 1$, which is equivalent to $p \leq q$,
or $\mathbb{P}(q|p) = 0$, which is equivalent to the orthogonality and mutual exclusivity 
of $p$ and $q$ (example 1).
The same holds, if $p$ and $q$ were elements of a Boolean algebra, where $p \leq q$
defines a logical relation and means that the proposition $p$ implies 
the proposition $q$. Therefore, $\mathbb{P}(q|p)$
can be considered an extension of this logical relation to the Hilbert space quantum logic.
This extended relation, however, 
is associated with a probability and introduces 
a continuum of new cases between the two classical cases
`$p$ implies $q$' and `$p$ rules out $q$'.

The transition probabilities introduced here include not only
the well-known quantum mechanical 
transition probabilities between pure states or wave functions, 
but further physically meaningful and experimentally verifiable novel cases 
where $\mathbb{P}(q|p)$ exists although $p$ is a projection on a subspace with
dimension higher than one (example 4). 
These novel cases 
become possible also in quantum logics 
that do not contain any projections on one-dimensional subspaces.
Such quantum logics are formed by the self-adjoint projections in the 
von Neumann algebras of the types II and III, 
which play an important role in quantum field theory 
and quantum statistical mechanics \cite{bratteli_rob}.

\bibliographystyle{abbrv}
\bibliography{Literatur}
\end{document}